\documentclass[aps, prl,twocolumn,showpacs]{revtex4}

\usepackage{graphicx}
\usepackage[ansinew]{inputenc}
\usepackage{array}
\usepackage{color}
\usepackage{amsmath}
\usepackage{amsxtra}
\usepackage{amstext}
\usepackage{amssymb}
\usepackage{latexsym}
\usepackage{dsfont}
\usepackage{verbatim}

\begin{document}

\title{Optical Control of a Quantum Rotor}
\author{L. F. Buchmann$^1$}
\author{H.~Jing$^2$}
\author{C. Raman$^3$}
\author{P. Meystre$^1$}
\affiliation{$^1$College of Optical Sciences and B2 Institute, University of Arizona, Tucson, Arizona 85721}
\affiliation{$^2$Department of Physics, Henan Normal University, Xinxiang 453007, P.R. China}
\affiliation{$^3$School of Physics, Georgia Institute of Technology, Atlanta, Georgia 30332, USA}
\date{\today}

\begin{abstract}
The possibility to coherently control a quantum rotor is investigated theoretically. The rotor is realized by an antiferromagnetic spin-1 Bose-Einstein condensate, trapped in the optical field of a Fabry-P\'erot resonator. By tuning the pumping field of the resonator, coherent control over the rotor is achieved. The technique is illustrated by the numerical simulation of a protocol that transforms the rotor's ground state into a squeezed state. The detection of the squeezed state via measurement of intensity-correlations of the cavity field is proposed. 
\end{abstract}

\maketitle
Among all quantum mechanical degrees of freedom, the spin stands out as a fundamental property with no classical analogue. While many effects of spin in crystals are well understood and readily seen in the macroscopic world through the magnetism of solids, the theoretical study of spin models continues to be a topic of active research. The realization of spinor Bose-Einstein condensates (BECs)~\cite{spinorBEC1}, with their remarkable control of dimensionality, external potentials and two-body interactions, has significantly broadened the experimental toolkit for the study of such models and already lead to a wide range of rich new physics, see~\cite{spinBECreview} for a recent review. 

Besides fundamental aspects, spinor BECs also hold promises for ultra-sensitive magnetometry with large spatial resolution~\cite{mukund}. In this context, much attention has been given to spin-squeezing~\cite{Vitigliano2011, Hamley2012, Bookjans2011}, a consequence of nonlinear spin-mixing dynamics. Some groups have also reported the observation of other quantum signatures in the population dynamics of such gases~\cite{Liu2009_2, Leslie2009, Klempt2010}.  However, in order to use spinor BECs as a nonclassical resource, coherent control over the quantum state has to be achieved, which is difficult due to the intrinsically nonlinear nature of the system. 

This letter proposes a scheme that allows coherent control over a quantum rotor, realized by an anti-ferromagnetic spin-1 BEC trapped within the field of an optical resonator~\cite{Barnett2010, Barnett2011, HuiJing2011}. We show how the optical field provides a convenient knob to coherently control the dynamics. As an example, we demonstrate the possibility to squeeze the quantum rotor by temporally modulating the pump-field of the optical cavity. Our results are more far-reaching however, as  they also provide a new mechanism to produce spin squeezing, since control over the rotor amounts to control over the spin degrees of freedom of the spinor gas. 

\begin{figure}
\includegraphics[width=0.8\columnwidth]{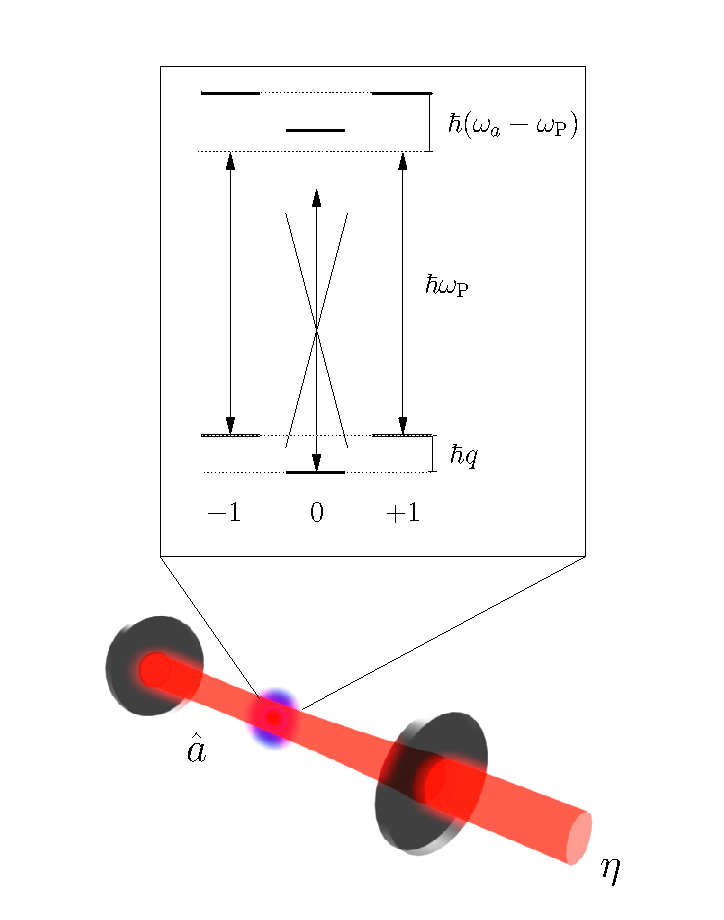}
\caption{A spin-1 BEC is trapped in an optical cavity and dispersively coupled to the linearly polarized cavity mode. The $m_f=0$ transition is forbidden by selection rules.}
\label{setup}
\end{figure}

We consider the spin-mixing dynamics of an anti-ferromagnetic spin-1 BEC in the case of a tightly trapped cloud, where all three spin populations can be described by a single spatial mode. The spin degree of freedom exhibits several peculiarities that make it particularly suitable for the preparation of non-classical states. Firstly, the internal time scales of spin systems are very slow compared to optical time scales. Secondly, the spin of a BEC is very well isolated from decoherence and thermal effects, with decoherence times in the range of seconds. This isolation is a natural requirement to perform coherent manipulations on the system. 

Before including the light field in our treatment, let us focus on the condensate alone. Its spin populations evolve due to spin-dependent collisions and the linear and quadratic Zeeman shifts. Since the two shifts commute and evolve on vastly different timescales, they can be studied independently. In our case, we focus on the quadratic shift alone. The Hamiltonian describing the system is 
\begin{equation}
H_\mathrm{col}=\frac{c_2}{N}\hat{F}^2-q\hat{b}_0^\dag\hat{b}_0,
\label{freeatomichamiltonian}
\end{equation}
where $c_2$ is the spin coupling, $N$ the total particle number and $\hat{F}=\hat{b}_i^\dag F_{ij}\hat{b}_j$,  with $F_{ij}$ spin-1 matrices and $\hat{b}_j$ bosonic annihilation operators for the three \hbox{$m_f=j\in \{\pm,0\}$} components. The second term is the quadratic Zeeman shift due to an external magnetic field, with $q$ being proportional to its squared absolute value.

It was shown in Ref.~\cite{Zhou2001} that the collision part of Hamiltonian (\ref{freeatomichamiltonian}) can be mapped onto an angular momentum Hamiltonian, $c_2{\bf \hat{L}}^2/(2N)$, acting on the Euclidean space spanned by $|x\rangle=(|+\rangle-|-\rangle)/\sqrt{2}$, $|y\rangle=(|+\rangle+|-\rangle)/(i\sqrt{2})$ and $|z\rangle=|0\rangle$, where $\{|\pm\rangle,|0\rangle\}$ is the spin basis. Physically, the orientation of the rotor gives the axis in real coordinate space along which the $z$-component of the spin vanishes. More recently it was shown how to extend that mapping to include the quadratic Zeeman shift~\cite{Barnett2010, Barnett2011}, resulting in an external potential $V(\hat \theta)$ 
\begin{subequations}\label{rotorhamiltonian}
\begin{eqnarray}
H_\mathrm{rot}&=&\frac{\hat{\bf L}^2}{2I}+qV(\hat{\theta}),\\
V(\hat{\theta})&=&\chi_1\sin^2(\hat{\theta})+\chi_2\sin^2(2\hat{\theta}).
\end{eqnarray}
\end{subequations}
Here $\chi_1=(N+3/2)$, $\chi_2=qN/(8c_2)$ and $I=N\hbar^2/c_2$ and functions of operators are defined by their power series. 
The ratio between the two terms in the potential is given by $\chi_1/\chi_2\approx q/(8c_2)$ and measures the balance between the Zeeman shift and the collision term. For an external field around 100 mG and a sodium condensate,  $\chi_1/\chi_2\approx 10^{-3}$, and we can neglect the term proportional to $\chi_2$, rendering the $V(\hat{\theta})$ proportional to $\sin^2(\hat{\theta})$. 

As a result of the symmetry of (\ref{freeatomichamiltonian}) under exchange of $m_f=\pm 1$ atoms,  the angular momentum along the z-axis, which in terms of atomic variables is given by $\hat{L}_z=\hbar(\hat{b}^\dag_+\hat{b}_+-\hat{b}^\dag_-\hat{b}_-)$, is a conserved quantity. Preparing the system initially with the same number of atoms in these two states reduces the number of degrees of freedom to only two. The two conjugate observables are given by the angle $\hat{\theta}$, measuring the population in the $m_f=0$ state, and the associated angular momentum $\hat{L}_\theta$, related to coherences between the $m_f=0$ and the other two spin components. Note that $\hat{\theta}$ is a periodic observable, but it is not related to the quantum phase \cite{Barnett2010}. 

We now show how it is possible to dynamically control the state of  the spinor gas and generate nonclassical rotor states by trapping it in an optical resonator driven by a laser of time-varying amplitude, see Fig.~\ref{setup}. If the laser field is linearly polarized along the direction of an external magnetic field and far detuned from the atomic transition,  the Hamiltonian for the compound system in the frame rotating at the laser frequency $\omega_l$ reads~\cite{HuiJing2011}
\begin{equation}
H_\mathrm{tot}=\hbar\Delta\hat{a}^\dag\hat{a}+\frac{\hat{L}_\theta^2}{2I}+\left(q+\hbar U_0\hat{a}^\dag\hat{a}\right) V(\hat{\theta})-i\hbar\eta(\hat{a}-\hat{a}^\dag)+\hat{H}_\kappa,
\label{hamiltonian}
\end{equation}
where the operator $\hat{a}$ annihilates a photon in the cavity mode and the atom-photon coupling is given by $U_0=g^2/(\omega_l-\omega_a)$ with $g$ the dipole coupling and $\omega_a$ the frequency of the atomic transition. We assume $U_0>0$ without loss of generality and also have introduced the pumping rate $\eta$, the cavity detuning $\Delta=\omega_c-\omega_l$ with $\omega_c$ the resonance frequency of the empty cavity and $H_\kappa$ accounting for optical dissipation with rate $\kappa$. Dissipation for the atomic system occurs on the timescale of seconds, while the dynamics considered here take place within milliseconds and we can safely neglect atomic dissipation. The evolution of the angular operator $\hat{\theta}$ is given by
\begin{eqnarray}
I\frac{\partial^2\hat{\theta}}{\partial t^2}&=&-2(q+\hbar U_0\hat{a}^\dag\hat{a})\nonumber\\
&\times&\left(\chi_1\sin(\hat{\theta})\cos(\hat{\theta})+2\chi_2\sin(2\hat{\theta})\cos(2\hat{\theta})\right)\label{thetaeq1}.
\end{eqnarray}
In the following we treat the driving field classically, and due to the slow evolution of the spin dynamics we can assume the optical field to follow the evolution of $\hat \theta$ adiabatically. Eliminating the optical field dynamics results in the effective potential
\begin{equation}
V_\mathrm{eff}(\hat{\theta})=V(\hat{\theta})+\frac{2\hbar\eta^2}{q\kappa}\tan^{-1}\left(\frac{2\Delta+2U_0V(\hat{\theta})}{\kappa}\right).
\label{effpot}
\end{equation}
and the effective Hamiltonian for the one-dimensional rotor becomes
\begin{align}
\mathcal{H}_\mathrm{eff}&=\frac{L_\theta^2}{2I}+qV_\mathrm{eff}(\theta).
\end{align}

Neglecting quantum fluctuations in the cavity field is not a priori unproblematic since for practical parameters the average number of photons in the cavity can be rather small, e.g. less than a thousand. The effective potential still provides a correct description for the rotor because the fluctuations in the cavity are correlated on a timescale of $\kappa^{-1}$, much faster than the characteristic timescales of the rotor. The rotor averages out quantum fluctuations. This is confirmed in our numerical calculations, where we have included fluctuations of the optical field. Before we present these results, however, let us discuss the merits of the optical field as a control tool to coherently manipulate the rotor. 

The qualitative nature of $V_{\rm eff}$ depends on the relations between $\Delta, U_0 N$ and $\kappa$. The inverse tangent is a monotonously increasing function that changes very slowly, except in an interval of order unity around the origin, where it increases steeply. Thus the minima and maxima of the original potential remain conserved. We can moreover expect that if the absolute value of the argument is very large, the qualitative distortion of the original potential is small. In terms of our parameters, such a regime is captured by the condition $|\Delta|\gg U_0N$, corresponding to the situation of a pump laser far detuned from the cavity resonance for any state of the rotor. In this case we can expand the potential around $U_0\langle V(\hat{\theta}) \rangle/\kappa$ and find after dropping constant terms
\begin{equation}
V_\mathrm{eff}(\hat{\theta})\approx \left(1+\frac{2\eta^2}{\kappa^2/4+\Delta^2}\right)V(\hat{\theta}),
\end{equation}
which is the trapping enhancement found in~\cite{HuiJing2011}.  

\begin{figure}
\includegraphics[width=0.9\columnwidth]{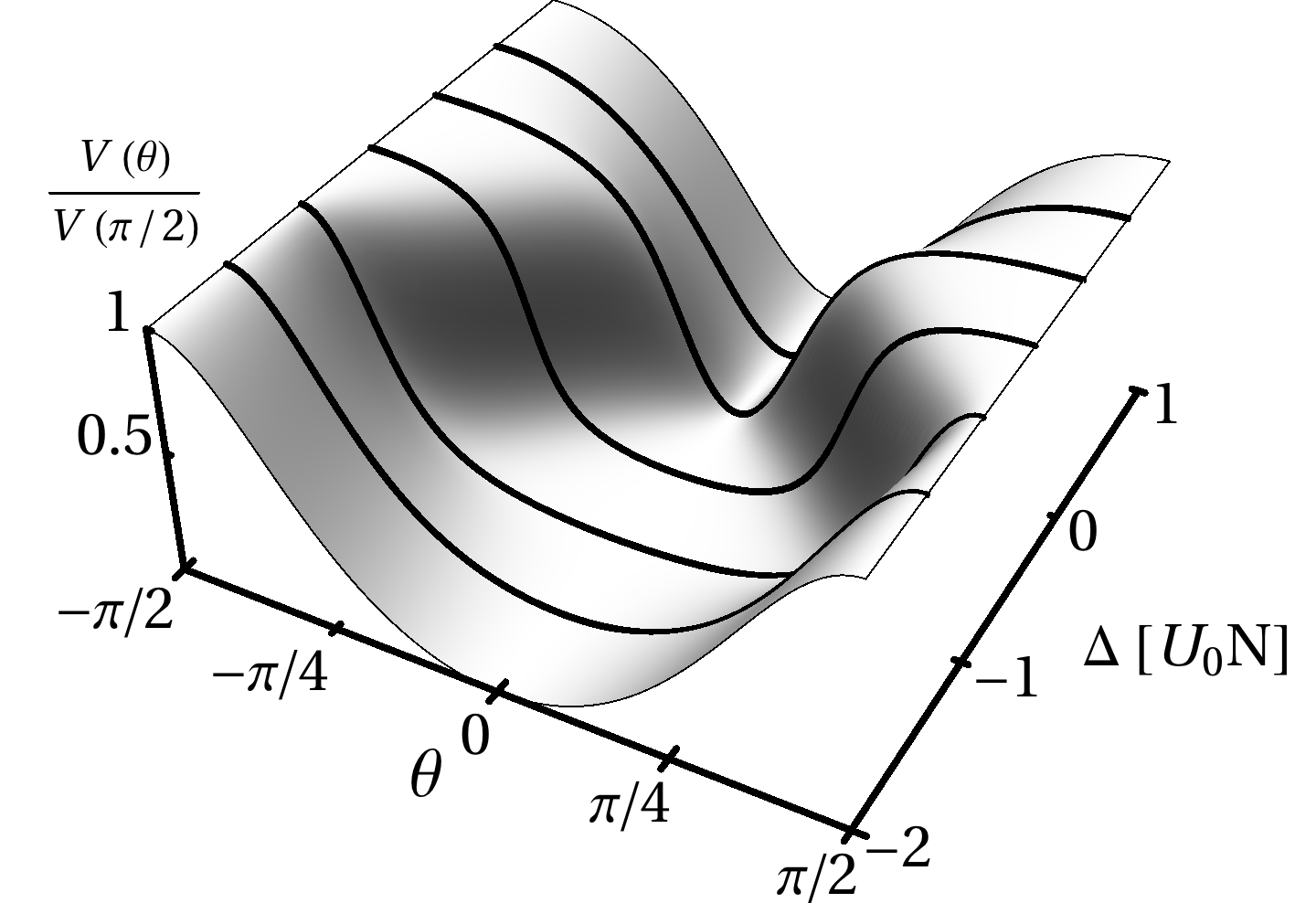}
\caption{Effective scaled rotor potential as a function of cavity-detuning and angle. See text for list of parameters.}
\label{potentials}
\end{figure}

For the resonator to significantly distort rather than merely scale the potential, two conditions need to be met: 
\begin{enumerate}
\item{The total range of detuning due to the rotors state, $U_0N/\kappa$, has to be much larger than unity.} 
\item{The argument of the inverse tangent has to cover regions around the origin, where the function changes characteristics.}
\end{enumerate}
The first condition requires a sufficiently strong atom-photon coupling, while the second condition can be satisfied by the right choice of laser frequency, i.e. $\Delta<\kappa$ and $|\Delta|<U_0N$. If both conditions are satisfied, the optical resonance will enhance the parts of $V(\hat{\theta})$ that fall on the steep portion of the inverse tangent around zero and flatten those beyond this region. For $\Delta=0$, the rotor's confinement is maximally tightened, whereas for $\Delta=-U_0N$  we are left with a predominantly flat effective potential. This behavior is illustrated in Fig.~\ref{potentials}, which shows the scaled potential as a function of $\theta$ and the pump-cavity detuning $\Delta$. 

The character of the effective potential will adapt to changes in $\eta$ and $\Delta$ on the fast time scale $\kappa^{-1}$, but we are free to change them as slowly as we want. This flexibility allows for efficient manipulations of the rotor's quantum state. For the remainder of this discussion, we will turn to the Schr\"odinger picture in $\theta$ space, which we also used for our numerical calculations. Let us assume for concreteness that the quantum rotor is initially in some state 
\begin{equation}
|\psi(\theta)\rangle=\sum_jc_j|\psi_j(\theta)\rangle,
\end{equation}
where $|\psi_j(\theta)\rangle$ are the eigenfunctions corresponding to a particular form of the effective potential. Under a  change of $V_\mathrm{eff}$ sufficiently slow that the adiabatic theorem holds, the coefficients $c_j$ do not change their distribution, but the eigenfunctions $\psi_j(\theta)$ become eigenfunctions $\psi_j'(\theta)$ of the new potential. This can be used to reset the system to its original parameters or to bring it to a desired regime of operation.  For a diabatic change, however, the new state is the projection of the old one on the new eigenfunctions, 
 \begin{equation}
|\psi'(\theta)\rangle=\sum_{i,j}c_j\langle\psi'_i(\theta)|\psi_j\rangle|\psi'_i\rangle. 
\end{equation}
Immediately following the switch the state still has the same wave function, but with a different distribution of eigenfunctions. Diabatic changes redistribute the weight of the eigenvalues of a wave function and hence the state's characteristics. 

With appropriate combinations of adiabatic and diabatic  transitions it is possible  to engineer desirable states, as we now illustrate on the concrete example of squeezed state generation. The idea is to diabatically switch between tight and wide potentials at the right instants in time and let the rotor evolve freely between the changes. This protocol was originally proposed to squeeze the motion of trapped ions~\cite{ions1}, but the fast switching times required make it challenging in such systems~\cite{ions2}. 
For typical experimental parameters the angular width of the ground state even in the shallow potential is very small, so that we can neglect the periodicity of the rotor and approximate both regimes by harmonic confinements with frequencies $\omega_1$ and $\omega_2$ for the tight and wide trap respectively -- these being easily found by an expansion of the potential. 

We assume that the rotor is initially in its ground state with respect to the trapping frequency $\omega_1$. The distributions of $\theta$ and $L_\theta$ are Gaussian with their respective zero-point widths. Suddenly changing the rotor's trapping frequency to the lower frequency $\omega_2$ leaves the wave-function unchanged; with respect to the new potential, however, its angular distribution is narrower than the zero-point width while it's angular momentum distribution is wider, i.e. the rotor is squeezed with squeezing angle 0 in the $\theta-L_\theta$ plane. In the new potential, the rotor undergoes its free time evolution and after a time $t=\pi/2\omega_2$ the angle is converted to angular momentum and vice versa. Now switching back diabatically to the initial, tight potential the situation is exactly reversed compared to the first change and the rotor experiences further squeezing, the squeezing angle now being $\pi/2$. After the time $t=\pi/2\omega_1$, this angle has changed to $\pi$ and we again switch to the wide potential. This process can be repeated and after each cycle, the uncertainty in one quadrature is scaled by the factor $(\omega_2/\omega_1)^2$, while the orthogonal quadrature is scaled with the inverse factor. This protocol can be understood as a discretized version of the squeezing Hamiltonian $\hat{a}^2+(\hat{a}^\dag)^2$ that is realized for example in parametric amplifiers. 

In practice, things differ of course from this idealized picture. First, it is impossible to perform truly instantaneous changes to the potential. Moreover, the timing of the switches is essential, but since typically the number of particles in the condensed gas is not constant during an experiment and also exhibits shot-to-shot fluctuations, the rotation frequencies of the rotor are not precisely known. A further limiting point is the anharmonic nature of the trapping potentials, which have a $\sin^2(\theta)$ dependence. These last two points become more relevant as more squeezing is achieved and the rotor's distribution becomes wider.

\begin{figure}
\includegraphics[width=0.6\columnwidth]{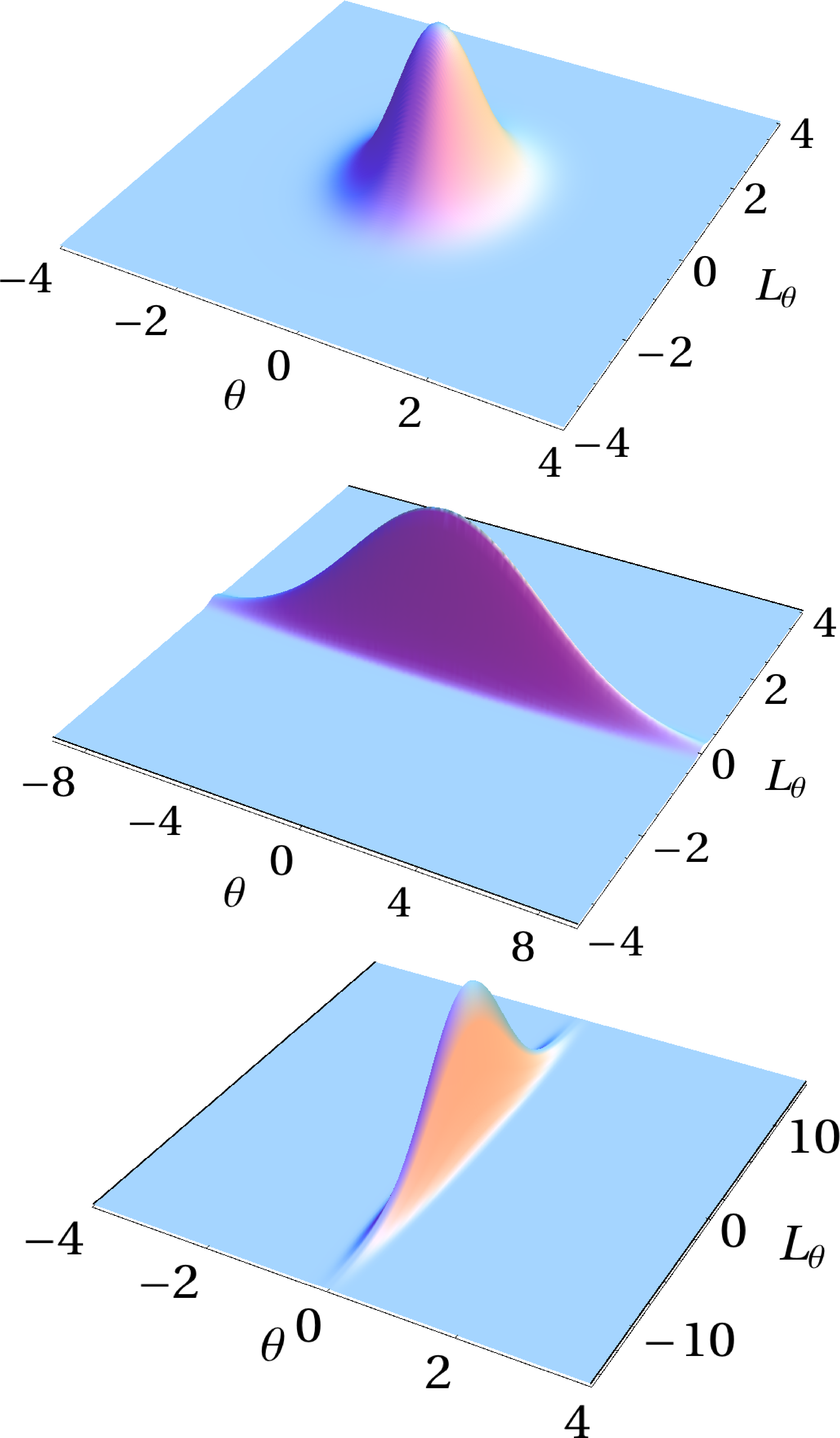}
\caption{Wigner distributions of the rotor.  (a) Ground-state distribution. (b) Distribution after the first switch of the trapping potential. (c) Distribution after completion of the squeezing cycle.  See text body for a list of system parameters. Units are the zero-point widths of the system.}
\label{wignerfun}
\end{figure}

To investigate these difficulties we have performed a numerical simulation of the protocol. Specifically, we solved the time dependent Schr\"odinger equation in $\theta$-space with a realistic time dependence of the effective potential, including the effects of photon noise through stochastic fluctuations of the effective potential and also chosen the atom number with a 5\% uncertainty to mimic potential experimental difficulties.  The result of these simulations is shown in Fig.~\ref{wignerfun} for a ${}^{23}\mathrm{Na}$ BEC consisting of $N\approx 10^4$  atoms with a number density of $5\times 10^{14}\mathrm{cm}^{-3}$. The pump laser is detuned from the sodium D1 line by $\Delta_a=2\pi\times1$~GHz and feeds a cavity with line-width $\kappa=2\pi\times 1$~MHz. Using these parameters, we find that the maximal confinement and the flattest effective potentials have harmonic frequencies of $\omega_1\approx 2\pi\times 43 \mathrm{kHz}$ and $\omega_2\approx 2\pi\times 7 \mathrm{kHz}$ respectively, separating the rotor and cavity timescales by more than a factor of 20. The rotor was initially in its ground state, see panel (a), and the squeezing after the potential switches is clearly visible.

There are several possibilities for the experimental characterization of the state of the rotor. The most straightforward approach is to perform a Stern-Gerlach type measurement, which is a projection on the number basis of the three components. While the angular distribution is directly related to the population of $m_f=0$ atoms, the rotor's angular momentum corresponds to coherences between $m_f=0$ and $m_f=\pm 1$ atoms and thus is not captures in this measurement. Furthermore, it is a destructive measurement technique, and thus not suitable to acquire large amounts of data. To fully resolve the rotor dynamics could involve dispersive imaging techniques that would give direct in-situ access to atomic coherences~\cite{rotorimag}. Continuous monitoring of the coherences, however, will generally require quantifying the measurement back-action on the rotor~\cite{Steve}. Moreover, additional imaging beams impose experimental difficulties. 

The necessary information is also contained in the output field of the optical resonator, provided that one accounts for its quantum nature. Assuming that the rotor does not change its state over the fast time scale $\kappa^{-1}$ we can expand the potential in Hamiltonian (\ref{hamiltonian}) as $V(\hat{\theta})=\langle V(\hat{\theta})\rangle+\hat{\delta V}$ and solve the equation of motion for the photon field operator to first order in $\hat{\delta V}$
\begin{align}
\hat{a}(t)=&\alpha_s+\hat{F}_\mathrm{in}+\gamma\hat{\delta V},
\label{expansion}
\end{align}
where $\alpha_s$ is the coherent steady-state amplitude, $\hat{F}_\mathrm{in}$ involves the vacuum input noise and
\begin{equation}
\gamma=i\eta U_0\left[\kappa/2+i(\Delta+U_0\langle V(\hat{\theta})\rangle)\right]^{-2}.
\end{equation}
Explicit expressions are easily found from Eq.~(\ref{hamiltonian}), but are omitted here for brevity. From these expressions one can find a non-vanishing contribution of $\langle\hat{\delta V^2}\rangle$ to the second order correlation function $g^{(2)}=\langle\hat{a}^\dag\hat{a}^\dag\hat{a}\hat{a}\rangle/[\langle{\hat{a}^\dag\hat{a}\rangle\langle{\hat{a}^\dag\hat{a}}\rangle}]$. Our numerical calculations indicate that after 5 squeezing cycles, which would take about 200 $\mu$sec,  the state of the rotor would lead to oscillations of $g^{(2)}$ with an amplitude of order $\sim 0.02$, which would render our proposal realizable in state-of-the-art experiments. 

In summary, we have investigated the possibility to control the quantum state of a spinor gas through an optical resonator. Using the rotor mapping for the quantum gas, we find the effective potential created by the optical field. Due to the slow spin-mixing dynamics, diabatic changes in the effective potential can be induced by tuning optical parameters. We demonstrated how such changes can be used to squeeze the rotor by means of a numerical simulation. Information about the rotor's state is imprinted on the intensity fluctuations of the cavity field and may  be used to verify squeezing. Further studies will involve the influence of in-situ measurements on the rotor, as well as the possibility to couple separate spinor clouds through the cavity field. 

LFB acknowledges helpful discussions with M.~Vengalattore. This work is supported by the DARPA ORCHID program through a grant from AFOSR, the U.S. National Science Foundation, and the U.S. Army Research Office.


\begin{thebibliography}{99}

\bibitem{spinorBEC1}D. M. Stamper-Kurn, M. R. Andrews, A. P. Chikkatur, S. Inouye, H.-J. Miesner, J. Stenger, W. Ketterle, Phys.~Rev.~Lett. {\bf 80}, 2027 (1998). 

\bibitem{spinBECreview}Y.~ Kawaguchi and Masahito Ueda, 10.1016/j.physrep.2012.07.005

\bibitem{mukund} M. Vengalattore, J. M. Higbie, S. R. Leslie, J. Guzman, L. E. Sadler, and D. M. Stamper-Kurn, Phys. Rev. Lett. {\bf 98}, 200801 (2007) 

\bibitem{Vitigliano2011} G.~Vitagliano, P.~Hyllus, I.~L.~Egusquiza and G.~T\"ath, Phys. Rev. Lett. {\bf 107}, 240502 (2011). 

\bibitem{Hamley2012} C.~D.~Hamley, C.~S.~Gerving, T.~M.~Hoang, E.~M.~Bookjans and M.~S.~Chapman, Nature Phys. {\bf 8}, 305 (2012).

\bibitem{Bookjans2011} E.~M.~Bookjans, C.~D.~Hamley, and M.~S.~Chapman, Phys. Rev. Lett. {\bf 107}, 210406 (2011).

\bibitem{Liu2009_2} Y.~Liu et al., Phys. Rev. Lett. {\bf 102}, 225301 (2009). 

\bibitem{Leslie2009} S.~R.~Leslie, J.~Guzman, M.~Vengalattore, Jay~D.~Sau, Marvin~L.~Cohen, and D.~M.~Stamper-Kurn, Phys. Rev. A {\bf 79}, 043631 (2009). 

\bibitem{Klempt2010} C. Klempt, O. Topic, G. Gebreyesus, M. Scherer, T. Henninger, P. Hyllus, W. Ertmer, L. Santos, and J. J. Arlt, Phys. Rev. Lett. {\bf 104}, 195303 (2010). 

\bibitem{Barnett2010} R.~Barnett, J.~D.~Sau and S.~Das Sarma, Phys. Rev. A {\bf 82}, 031602(R) (2010). 

\bibitem{Barnett2011} R.~Barnett, H.~-Y.~Hui, C.~-H.~Lin, J.~D.~Sau and S.~Das Sarma, Phys. Rev. A {\bf 83}, 023613 (2011). 

\bibitem{HuiJing2011} H.~Jing, D.~S.~Goldbaum, L.~Buchmann and P.~Meystre, Phys. Rev. Lett. {\bf 106}, 223601 (2011). 

\bibitem{Zhou2001} F.~Zhou, Phys. Rev. Lett. {\bf 87}, 080401 (2001).

\bibitem{ions1} D.~J.~Heinzen and D.~J.~Wineland, Phys.~Rev.~A {\bf 42}, 2977 (1990). 

\bibitem{ions2} J Alonso, F M Leupold, B C Keitch and J P Home, ArXiv:1208.3986 (2012). 

\bibitem{rotorimag} I.~Carusotto and E.~J.~Mueller, J. Phys. B: At. Mol. Opt. Phys. {\bf 37} S115, (2004). 

\bibitem{Steve} S.~K.~Steinke et al.  arXiv:1211.2870 (2012).

\end{thebibliography}
\end{document}